\documentclass[12pt,a4paper,superscriptaddress,preprint]{revtex4}
\usepackage[dvips]{graphicx}
\usepackage{amssymb,amsmath}
\usepackage{bm}
\usepackage{color}

\begin{document}

\title{Voter model on the two-clique graph}
\date{\today}
\author{Naoki Masuda}
\affiliation{Department of Engineering Mathematics,
Merchant Venturers Building, University of Bristol,
Woodland Road, Clifton, Bristol BS8 1UB, United Kingdom}
\affiliation{CREST, JST, 4-1-8, Honcho, Kawaguchi, Saitama 332-0012, Japan}
\email{naoki.masuda@bristol.ac.uk}

\begin{abstract}
I examine the mean consensus time (i.e., exit time) of the voter model in the so-called two-clique graph. The two-clique graph is composed of two cliques interconnected by some links and considered as a toy model of networks with community structure or multilayer networks. I analytically show that, as the number of interclique links per node is varied, the mean consensus time experiences a crossover between a fast consensus regime [i.e., $O(N)$] and a slow consensus regime [i.e., $O(N^2)$], where $N$ is the number of nodes. The fast regime is consistent with the result for homogeneous well-mixed graphs such as the complete graph. The slow regime appears only when the entire network has $O(1)$ interclique links. The present results suggest that the effect of community structure on the consensus time of the voter model is fairly limited.
\end{abstract}

\maketitle

\section{Introduction}

In collective opinion formation taking place in a population of interacting agents, competing opinions are often approximately as strong as each other. The voter model
is a simple stochastic process to represent competitive
dynamics between equally strong states (i.e., opinions)
\cite{Liggett1985book,Redner2001book,Barrat2008book,Castellano2009RMP,Krapivsky2010book,SenChakrabarti2013book}.
In the voter model, an
agent flips its state to a new state at a rate proportional to the number of
neighboring agents that possess the new state.
In a finite connected network, consensus of one state is always 
the eventual outcome of the voter model dynamics.

Social networks in which opinion formation takes place are usually
complex. In particular, community structure, in which connection is
dense within groups and sparse across different groups, is a
hallmark of a majority of social networks. A community would correspond to
a circle of friends, school class, organization, household, and so on
\cite{Fortunato2010PhysRep}.
Consensus formation in networks with community structure may need a long time
because communities are sparsely connected to each other by
definition and different communities have to align their states for the consensus in the entire network to be reached. In fact, in the voter
model in metapopulation networks in which agents randomly
diffuse from one metapopulation to another,
a small diffusion rate (corresponding to sparse
connectivity between communities) slows down consensus
\cite{Baronchelli2009JSM}.
In addition, consensus is often hampered in other opinion formation models
when the network possesses community structure
\cite{Lambiotte2007PhysRevE_majority,Lambiotte2007JStatMech,Toivonen2009PRE,Candia2008JStatMech,WangWuWangFu2008PhysRevE,DasguptaPanSinha2009PhysRevE}.

However, the extent to which the community structure slows down the consensus dynamics is unclear. Previous numerical results suggest
that the dependence of the time to consensus on the number of nodes does not differ between networks with and without community structure \cite{Castello2007EPL}. The consensus time is also independent of the network structure for a related model of language exchange \cite{Baxter2008PRL}.
In the present study, I confine myself to a toy network model mimicking community structure and also multilayer networks \cite{Kivela2013arxiv}, called the two-clique graph. The voter model in this graph 
was briefly analyzed in Ref.~\cite{Sood2008PRE}. For the two-clique graph,
I reveal the scaling relationship between the time to consensus and the number of nodes, which depends on the number of links connecting two cliques.

\section{Model}\label{sec:model}

Consider a graph in which
each of the two cliques has $N$ nodes
\cite{Sood2008PRE}; the entire network has $2N$ nodes.
The two cliques are connected by $M$ ($0< M\le N^2$) 
interclique links. Each node has
\begin{equation}
C\equiv M/N
\end{equation}
interclique links on average. The interclique links are either
regularly placed such that each node has (approximately) $C$ interclique links or randomly placed such that the number of interclique links that a node possesses obeys a binomial distribution with mean $C$. 

I run a variant of the two-state voter model according to the link dynamics rule
\cite{Castellano2005AIPConf,Suchecki2005EPL,Antal2006PRL,Sood2008PRE}
on this network. Specifically, each node takes either of the two states
$\mathbf{0}$ and $\mathbf{1}$. Initially,
$N/2$
voters in each clique are assumed to be in the $\mathbf{0}$ state. The other $N/2$ voters in each clique are in the $\mathbf{1}$ state. In each
time step, I randomly pick a link with the equal probability, i.e., 
$1/[N(N-1)+M]$, and then select one of the two endpoints of the link with probability $1/2$. Then, the selected node copies the state of the other endpoint of the link.
Then, I move forward the clock by time $1/2N$ such that each node is updated once on average per unit time. The dynamics eventually reaches the consensus of either state. Denote the
consensus time and its mean by $T$ and $\left<T\right>$, respectively.

\section{Fokker-Planck equation}\label{sec:FP}

The Fokker-Planck equation for this dynamics
was previously formulated
\cite{Sood2008PRE}.
For the Fokker-Planck equation to be valid, it is necessary that
each node has exactly $C$ interclique links or 
$C$ is large such that the fluctuation in the number of interclique links per node is negligible.
The Fokker-Planck
equation in terms of the density of $\mathbf{1}$ voters in the two cliques, denoted by
$\rho_1$ and $\rho_2$, is given by
\begin{align}
\frac{\partial P}{\partial t}=&
-\frac{\partial}{\partial \rho_1}\left[
\frac{C}{N+C}\left(\rho_2-\rho_1\right)P\right]
-\frac{\partial}{\partial \rho_2}\left[
\frac{C}{N+C}\left(\rho_1-\rho_2\right)P\right]\notag\\
&+\frac{1}{2}\frac{\partial^2}{\partial^2 \rho_1^2}
\left[\left(\frac{\rho_1\left(1-\rho_1\right)}{N+C}+
\frac{\frac{C}{2N}\left(\rho_1+\rho_2-2\rho_1\rho_2\right)}{N+C}
\right)P\right]\notag\\
&+\frac{1}{2}\frac{\partial^2}{\partial^2 \rho_2^2}
\left[\left(\frac{\rho_2\left(1-\rho_2\right)}{N+C}+
\frac{\frac{C}{2N}\left(\rho_1+\rho_2-2\rho_1\rho_2\right)}{N+C}
\right)P\right],
\label{eq:FP}
\end{align}
where $P=P(\rho_1, \rho_2, t)$ represents the probability density.

When $C\gg 1$, Eq.~\eqref{eq:FP} implies that
the drift term dominates over the diffusion term.
This case was previously solved for more general network structure
by adiabatic approximation \cite{Baxter2008PRL,Constable2014arxiv}.
In the case of the 
two-clique graph, the density in both cliques
relaxes to $(\rho_1 + \rho_2)/2$ on a fast time scale.
The dynamics on a slow time scale, which leads to the consensus in the entire population, is essentially the same as that for the complete graph.
Therefore, $\left<T\right>\approx 2N\ln 2$ \cite{BenAvraham1990JPhysA}.

When $C=O(1)$, the drift and diffusion terms are comparable.
In this case, 
the problem is essentially two-dimensional and seems
difficult to solve.

When $C\ll 1$, the diffusion terms are dominant
on a fast time scale. In this situation, the approximate
consensus within each clique may be reached before the two cliques effectively start to interact.
If this is the case, the bias terms play a role after the consensus
in each clique has been reached. 
In fact, the Fokker-Planck approximation given by Eq.~\eqref{eq:FP}
breaks down
when $C\ll 1$ because the number of interclique links crucially 
differs node by node.
In other words, most nodes possess either zero or one interclique link, and 
not having any interclique link and having one interclique link may
result in substantially different behavior of nodes.

\section{Mean consensus time obtained from the coalescing random
  walk}\label{sec:coalescing RW}

In this section, I
theoretically determine the dependence of $\left<T\right>$ on $N$
using a random walk method.
The same scaling results as those derived in this section can be obtained with the use of a different, more intuitive, analysis method (Appendix~\ref{sec:1 shortcut}). However, the method shown in Appendix~\ref{sec:1 shortcut} is valid only when $C\ll 1/N$, which is unrealistic. The analysis in this section is valid for the entire range of $C$.

The dual process of the voter model is the coalescing random walk,
in which walkers visiting the same node coalesce into a single walker. For arbitrary networks, the consensus time is equal to the time needed for the $N$ simple random walkers, one walker initially located per node, to coalesce into one 
\cite{Liggett1985book,Donnelly1983MPCPS,Durrett1988book,Cox1989AnnProb}. The time
needed for the last two walkers to coalesce is considered to
dominate the consensus time. Therefore, in this section I estimate the mean consensus time by analyzing the mean time before the two walkers starting from random positions meet.

\subsection{When each node has many interclique links}\label{sub:dual large C}

When $C\gg 1$, the number of interclique links for a node does not differ much among the nodes. Therefore, I assume that all nodes in the same clique are structurally the same, as implicitly assumed in the derivation of the Fokker-Planck equation [Eq.~\eqref{eq:FP}]. The following analysis is also valid when each node has exactly $C$ interclique links and $C=O(1)$.

Denote by $p_1(t)$ the probability that the two walkers
are located at different nodes in the same clique at time $t$.
Denote by $p_2(t)$ the
probability that the two walkers are located in the opposite cliques
at time $t$. Finally, $r(t)$  
is the probability that the two walkers coalesce at time
$t$. In each time step, one of the two walkers is selected with probability $1/2$ and
moves to a neighbor according to the simple random walk. The selected walker moves to a neighbor with probability $1/(N+C-1)\equiv 1/\Delta$. The network under consideration is regular. Therefore, the simple random walk is equivalent to selecting an arbitrary link and its direction with probability $1/2M$ and moving a walker (if any) along the direction of the selected link, up to a time rescaling.

I obtain
\begin{equation}
\left(p_1(t)\atop p_2(t)\right) = A^t \left(p_1(0)\atop p_2(0)\right)
\end{equation}
and
\begin{equation}
r(t+1) = v_1 p_1(t) + v_2 p_2(t),
\end{equation}
where
\begin{equation}
A\equiv \frac{1}{\Delta}\begin{pmatrix}
N-2 & \frac{(N-1)C}{N}\\
C & N-1
\end{pmatrix}
\end{equation}
and
\begin{equation}
\bm v \equiv (v_1\; v_2) = \frac{1}{\Delta}\left(1\quad \frac{C}{N}\right).
\end{equation}

By using
\begin{equation}
(I-A)^{-1} = \frac{\Delta N}{(N+C)C} 
\begin{pmatrix}
C & \frac{(N-1)C}{N}\\
C & C+1
\end{pmatrix},
\end{equation}
one can verify
\begin{align}
\sum_{t=1}^{\infty} r(t) =& \sum_{t=1}^{\infty}
\bm v A^{t-1} \left(p_1(0) \atop p_2(0)\right)\notag\\
=& \bm v (I-A)^{-1} \left(p_1(0) \atop p_2(0)\right)\notag\\
=& p_1(0) + p_2(0) = 1
\end{align}
regardless of $p_1(0)$ and $p_2(0)$. By ignoring the transient process in which the $N$ random walkers coalesce into two random walkers, the
mean consensus time is evaluated as
\begin{align}
\left<T\right>\approx \sum_{t=1}^{\infty}t r(t) =& 
\bm v (I-A)^{-2}
\left(p_1(0) \atop p_2(0)\right)\notag\\
=& (1\; 1)(I-A)^{-1} \left(p_1(0) \atop p_2(0)\right)\notag\\
=& \frac{(N+C-1)N}{(N+C)C}
\left[2C p_1(0) + \left(1-\frac{C}{N}+2C\right)p_2(0)\right].
\label{eq:<T> via coalescing RW large C}
\end{align}

If $C\gg 1$, I obtain $2C p_1(0)\gg (1-C/N)p_2(0)$ unless $p_1(0)=0$, which leads to $\left<T\right>=O(N)$. It should be noted that, even for $C=O(1)$, Eq.~\eqref{eq:<T> via coalescing RW large C} implies
$\left<T\right>=O(N)$.

\subsection{When each node has at most one interclique link}\label{sub:dual small C}

Equation~\eqref{eq:<T> via coalescing RW large C} extrapolated to the case
$C<1$ indicates
$\left<T\right>=O(N/C)$. In particular, substitution of $C=O(1/N)$
in Eq.~\eqref{eq:<T> via coalescing RW large C} yields
$\left<T\right>=O(N^2)$. However, the assumption that each node has $C$ interclique links, which justified the annealed approximation (i.e., each node has exactly $C$ interclique links even if $C$ is not integer) developed in Sec.~\ref{sub:dual large C}, breaks down when $C< 1$. When $C<1$, some nodes do not possess any interclique link, whereas other nodes typically possess one interclique link. A single-step random walk starting from a node without an interclique link and that with an interclique link may be substantially different because only the latter allows the transition to the opposite clique.

In this section, I carry out a quenched
analysis (i.e., number of interclique links that each node possesses is explicitly considered) of the coalescing random walk for the case $C<1$.
Assume that $NC$ nodes in each clique has one interclique link each, and $N(1-C)$ nodes in each clique does not have any interclique link.
I have implicitly assumed that
different interclique links do not share an
endpoint. 
These assumptions exactly hold true when $C\ll 1$.
I consider two coalescing random walkers starting from different positions and assess the coalescing time, as was done in Sec.~\ref{sub:dual large C}.

\begin{figure}
\begin{center}
\includegraphics[width=8cm]{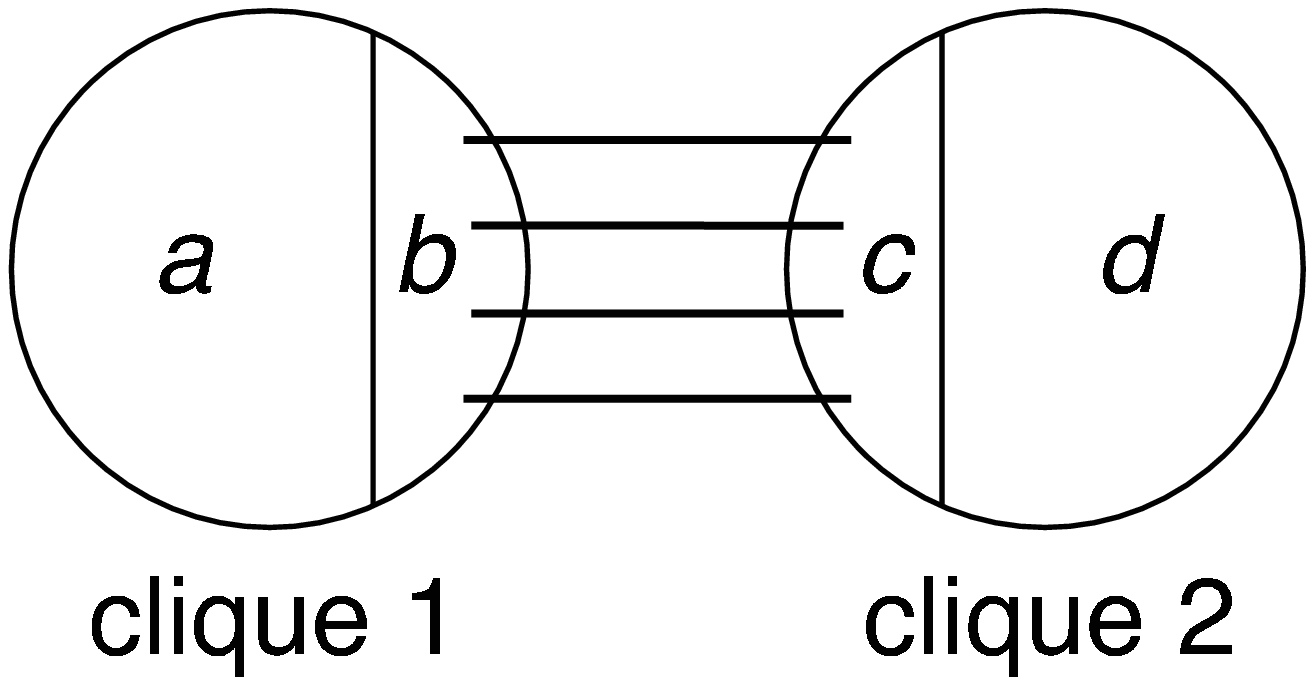}
\caption{Schematic of the four classes of nodes when $C<1$.}
\label{fig:4 classes of nodes}
\end{center}
\end{figure}

The nodes in the two-clique graph are divided into four 
equivalent classes, as shown in Fig.~\ref{fig:4 classes of nodes}.
The first class of nodes, which is called class $a$, contains $N(1-C)$ nodes in clique 1 that are not an endpoint of any interclique link. The second class, which is called class $b$, contains $NC$ nodes in clique 1 that are an endpoint of an interclique link. The third class, which is called class $c$, contains $NC$ nodes in clique 2 that are an endpoint of an interclique link. The fourth class, which is called class $d$, contains $N(1-C)$ nodes in clique 2 that are not an endpoint of any interclique link.

\begin{figure}
\begin{center}
\includegraphics[width=10cm]{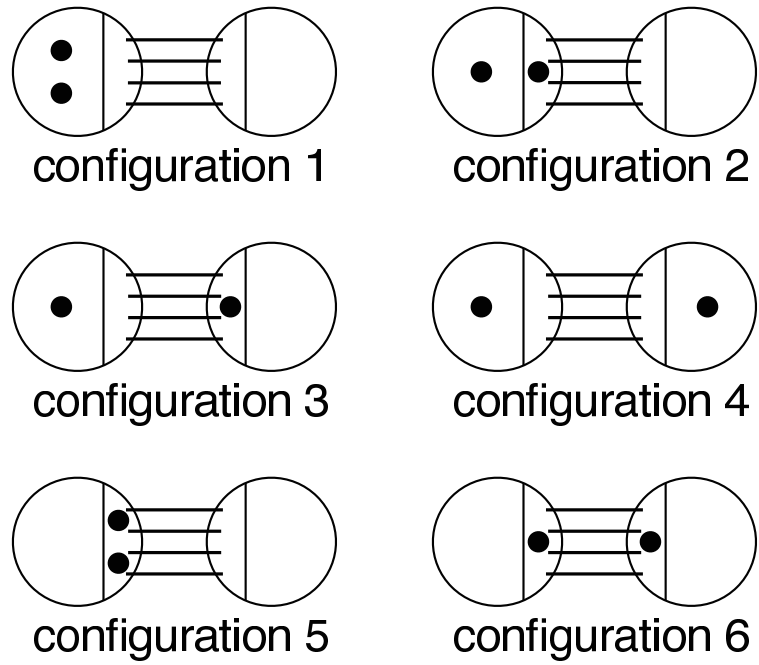}
\caption{Schematic of the six configurations of the coalescing random walk when $C<1$. It should be noted that clique 1 may correspond to either the left or right clique in the figure.}
\label{fig:6 configurations}
\end{center}
\end{figure}

At any time $t$, the coalescing random walk takes either of the following six configurations, as shown in Fig.~\ref{fig:6 configurations}, unless the two walkers coalesce. Denote by $(i,j)$ ($i,j\in \{a, b, c, d\}$) the event that two random walkers are located at a class $i$ node and a class $j$ node.
Denote by $p_1(t)$ the probability that both walkers visit
two different nodes in class $a$ (i.e., $(a, a)$)
or in class $d$ (i.e., $(d, d)$) at time $t$.
The sum of the probability of $(a, b)$ and that of $(c, d)$ is denoted by
$p_2(t)$. The sum of the probability of $(a, c)$ and that of $(b, d)$ is denoted by $p_3(t)$. The probability of $(a, d)$ is denoted by $p_4(t)$. The sum of the probability of $(b, b)$ and that of $(c, c)$ is denoted by $p_5(t)$. Finally, the probability of $(b, c)$ is denoted by $p_6(t)$.

In each time step, one of the $N(N+C-1)$ links in the network is selected with the equal probability, i.e., $1/[N(N+C-1)]$. Then, one of the two endpoints of the link selected with probability $1/2$ adopts the state of the other endpoint.
To explain the calculation of $A_{ij}$, i.e., the transition probability from configuration $j$ to configuration $i$ ($1\le i, j\le 6$),
consider configuration 1, in which the two walkers occupy different class-$a$ nodes (Fig.~\ref{fig:6 configurations}). There are three possible types of transition in one time step. First, one of the two walkers moves to a node in class $b$. This event occurs with probability
$A_{21} = 2NC/[N(N+C-1)]\times (1/2) = C/(N+C-1)$. Second, the two walkers coalesce with probability $2/[N(N+C-1)]\times (1/2) = 1/[N(N+C-1)]$. Otherwise, the configuration does not change such that $A_{11}=1- C/(N+C-1) - 1/[N(N+C-1)]$.

One can write down $A=(A_{ij})$, the six-dimensional vector $\bm v$, and the mean coalescing time as a linear sum of $p_1(0)$, $\ldots$, $p_6(0)$ in the same manner as in Sec.~\ref{sub:dual large C}. The detailed calculations are shown in Appendix~\ref{sec:calculation 6 configurations}.
In summary, the obtained scaling reads
\begin{equation}
\left<T\right> = O(N/C).
\label{eq:<T> via coalescing RW small C}
\end{equation}

Substitution of $C=O(1/N)$
in Eq.~\eqref{eq:<T> via coalescing RW small C} yields
$\left<T\right>=O(N^2)$.
This is the same scaling as the case of the one-dimensional lattice \cite{Cox1989AnnProb,Castellano2009RMP,Krapivsky2010book}.
It should be noted that, even if $C=O(1/N)$, the network diameter is equal to just 3, and the mean path length between a pair of nodes is small; it is approximately equal to $(1/2)\times 1 + (1/2)\times 3 = 2$ independent of $N$.

\section{Numerical simulations}\label{sec:numerical}

I perform $10^3$ runs of the voter model
for a given set of parameter values
$(N, M)$ and calculate $\left<T\right>$. A different two-clique
graph is generated for each run.
The initial condition is such that half the nodes randomly selected from each clique takes the $\mathbf{0}$ state and the other half the $\mathbf{1}$ state.
Other details of the numerical procedure are provided in Sec.~\ref{sec:model}.

\begin{figure}
\begin{center}
\includegraphics[width=10cm]{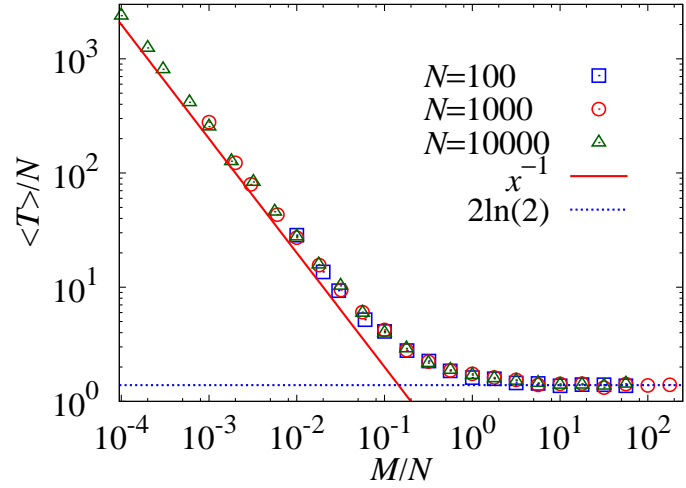}
\caption{(Color online) Relationship between the normalized mean consensus time,
$\left<T\right>/N$, and the number of interclique links per node, $M/N$. The solid line represents the relationship $\left<T\right>/N \propto \left(M/N\right)^{-1}$ as guides to the eye.}
\label{fig:scaling}
\end{center}
\end{figure}

By factoring out $1/(N+C)$ on the right-hand side of Eq.~\eqref{eq:FP}
and assuming $C\ll N$, I obtain the the following scaling ansatz:
\begin{equation}
\frac{\left<T\right>}{N} = f\left(\frac{M}{N}\right),
\label{eq:scaling ansatz}
\end{equation}
where $f$ is a scaling function.
In Fig.~\ref{fig:scaling},
$\left<T\right>/N$ is plotted against
$M/N$ for $N=10^2$, $10^3$, and $10^4$ and various $M$ values. The results for
different values of $N$ and $M$
collapse on a single curve, confirming the validity of Eq.~\eqref{eq:scaling ansatz}.

When $C=M/N\gg 1$, the network approaches the complete
graph such that $\left<T\right>=2N\ln 2$. In fact,
$\left<T\right>\approx 2N\ln 2$ holds true even if $C=O(1)$;
the horizontal dotted line in Fig.~\ref{fig:scaling}
indicates $\left<T\right>/N
=2\ln 2$. When $C=O(1/N)$, combination of
Eqs.~\eqref{eq:<T> via coalescing RW small C} and \eqref{eq:scaling ansatz} yields $f(x)\propto x^{-1}$ as $x\to 0$.
Figure~\ref{fig:scaling} indicates that this relationship holds true for small $x$;
the solid line represents $\left<T\right>/N \propto \left(M/N\right)^{-1}$.

\section{Discussion}

I examined the consensus time of a variant of the voter model on
the two-clique graph. Theoretically, 
the mean consensus time $\left<T\right> = O(N)$ when there are 
many (i.e., $\gg 1$) interclique links per node. Numerically, $O(1)$ interclique links per node is sufficient to realize the same scaling. When the number of interclique links per node is much smaller than unity, $\left<T\right> = O(N^2)$. The crossover between the two regimes seems to occur at approximately one interclique link per node (Fig.~\ref{fig:scaling}). Therefore, the voter model dynamics is considerably decelerated only when the two cliques are very sparsely connected. 
It is straightforward to extend the present results to the case of
more than two cliques.

The present results are consistent
with the previous numerical results showing that networks with
community structure in which intercommunity links are not rare
yield $\left<T\right>=O(N)$
\cite{Castello2007EPL}. A social network with an extremely sparse
connectivity between communities is unrealistic. It may bear some
realism in the context of genetic evolutionary dynamics, for which
invasion dynamics between sparsely interacting populations was
recently analyzed \cite{Altrock2011PlosComputBiol}.

In general, the two-clique graph defined in the present study is not regular in the node degree. In nonregular networks, behavior of the voter model depends on the rule according to which the node's state is updated
\cite{Castellano2005AIPConf,Suchecki2005EPL,Antal2006PRL,Sood2008PRE,Barrat2008book}.
The so-called link dynamics rule was used in the present study. In general, the results remain the same under different updating rules (invasion process and the so-called voter model rule) if the network is regular.
The Fokker-Planck equation (Sec.~\ref{sec:FP}) and the dual process (Sec.~\ref{sec:coalescing RW}) were implicitly considered on regular networks. In fact, the degree of a node in the two-clique graph is
equal to $N-1$ plus the number of interclique links. When the interclique links are placed randomly, the number of interclique links obeys the binomial distribution. However, its mean (i.e., $C$) and standard deviation are much smaller than $N-1$ in the parameter range of interest (i.e., $C=O(1)$ or smaller), rendering the network approximately regular. Therefore, the present results are considered to be robust with respect to the updating rule.

The two-clique graph can be regarded as a simple multilayer network with two layers \cite{Kivela2013arxiv}. In this context, the
Laplacian spectrum of multilayer networks
in which each layer is a general network is a useful tool \cite{Gomez2013PRL,Radicchi2013NatPhys}. Analyzing the current model and its extensions under the framework of multilayer networks warrants future work. 

\section*{Acknowledgments}

I thank Sidney Redner for valuable discussion throughout the current work.
I also thank Yuni Iwamasa and Taro Takaguchi for careful reading of the manuscript. N.M. acknowledges the support provided through CREST, JST.

\appendix

\section{Assessing the mean consensus time when $C\ll 1$}\label{sec:1 shortcut}

In this section, I estimate
$\left<T\right>$ in a hypothetical situation in which
consensus in each clique
is realized fast enough before an interclique link is selected to trigger interaction of the two cliques.
If the two cliques are disconnected,
the consensus in each clique is reached with mean time $N\ln 2$.
The mean time before an interclique link is selected, denoted by
$\left< t_{\rm ic}\right>$, is given by
\begin{equation}
\left< t_{\rm ic}\right> = \frac{1}{N}\left[\frac{CN}{N(N-1)+CN}\right]^{-1} = O(1/C),
\end{equation}
because
there are $CN$ interclique links and $N(N-1)$ intraclique links, and selection of a link consumes time $1/N$.
Therefore, the condition under which the following adiabatic approximation is valid is given by $N\ln 2 \ll O(1/C)$, i.e., $C\ll 1/N$.

Because links are implicitly assumed to be unweighted, $C\ge 1/N$, where the equality is realized when there is just one interclique link in the entire network. Therefore, the condition $C\ll 1/N$ is never satisfied. Nevertheless, the arguments in the remainder of this section turn out to predict the correct dependence of $\left<T\right>$ on $C$ when $C (\ge 1/N)$ is small, which was derived in Sec.~\ref{sub:dual small C}. If weighted links are allowed, $C\ll 1/N$ can be realized if, for example, there are $O(1)$ interclique links whose weights are much smaller than unity.

Under the assumption $C\ll 1/N$, the consensus is reached in cliques 1 and 2 on a fast time scale.
The consensus within each clique implies
the consensus of the entire network with probability $1/2$.
Otherwise, I assume without loss of generality that 
clique 1 reaches the $\mathbf{0}$ consensus and clique 2
reaches the $\mathbf{1}$ consensus. This event
occurs with probability $1/2$.
In the latter case, the consensus of the entire network occurs on a slow time scale. 

If the two cliques have reached the consensus of the opposite states,
without loss of generality, the event that happens next is
invasion of the $\mathbf{0}$ state into a node in clique 2
via an interclique link. This event occurs when an interclique link is selected for an update, which takes mean time $\left<t_{\rm ic}\right>$.

Then, one of the following two scenarios ensues.
In the first scenario,
state $\mathbf{0}$ fixates in clique 2, and the consensus of the entire network is reached. This event occurs with probability $1/N$ \cite{Nowak2006book,Ewens2010book}.
Under the condition that state $\mathbf{0}$ fixates in clique 2,
the mean fixation time in clique 2 is equal to (e.g., \cite{Sood2008PRE,Ewens2010book})
\begin{equation}
N\frac{1-\frac{1}{N}}{\frac{1}{N}}\ln \frac{1}{1-\frac{1}{N}}=O(N).
\label{eq:<T> when invasion is successful}
\end{equation}
Therefore, the consensus time is dominated by $\left<t_{\rm ic}\right>=O(1/C)$.
Here I ignored the contribution of $N\ln 2$ derived from the initial
intraclique consensus and that given by
Eq.~\eqref{eq:<T> when invasion is successful}
because $C\ll 1/N$ is assumed.

In the second scenario, state
$\mathbf{1}$ fixates in clique 2. This scenario occurs with probability
  $(N-1)/N$. Under the condition that state
$\mathbf{1}$ fixates in clique 2,
the mean fixation time in clique 2 is equal to 
\begin{equation}
N\frac{\frac{1}{N}}{1-\frac{1}{N}}\ln \frac{1}{\frac{1}{N}}=O(\ln N).
\end{equation}
In this case, the situation in which cliques 1 and 2
are in the consensus of the opposite states is revisited.
Then, either state invades the other state in the opposite clique
after mean time $\left<t_{\rm ic}\right>$,
and the same process repeats until the consensus of the
entire network is reached.

Therefore, I obtain
\begin{align}
\left<T\right>\approx & N\ln 2 + \frac{1}{2}\frac{1}{N}
\left<t_{\rm ic}\right> + \frac{1}{2}\frac{N-1}{N}\frac{1}{N}
\left[\left<t_{\rm ic}\right>+O(\ln N)+\left<t_{\rm ic}\right>\right]\notag\\
+& \frac{1}{2}\left(\frac{N-1}{N}\right)^2\frac{1}{N}
\left[\left<t_{\rm ic}\right>+O(\ln N) + \left<t_{\rm ic}\right>+O(\ln N) + \left<t_{\rm ic}\right>\right] + \cdots \notag\\
=& N\left<t_{\rm ic}\right> = O(N/C).
\label{eq:<T> for small M}
\end{align}
Equation~\eqref{eq:<T> for small M} implies that consensus is much slower as compared to the case of the complete graph [i.e., $\left<T\right>=O(N)$].
Extrapolation of Eq.~\eqref{eq:<T> for small M} to the case of
$O(1)$ interclique links in the networks, i.e., $C=O(1/N)$,
would lead to $\left<T\right>=O(N^2)$, which is actually correct as theoretically and numerically shown in Secs.~\ref{sec:coalescing RW} and \ref{sec:numerical}, respectively. 

\begin{figure}
\begin{center}
\includegraphics[width=10cm]{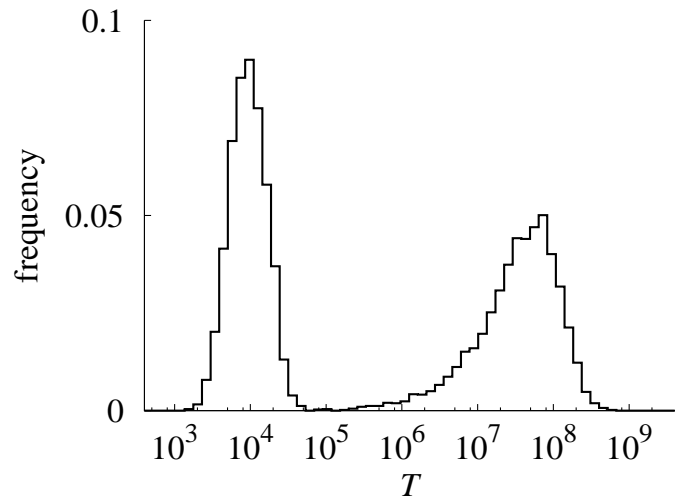}
\caption{Distribution of $T$ on the basis of $10^4$ runs. I set
$N=10^4$ and $M=1$. The vertical axis represents the fraction of runs that fall in the time window specified on the horizontal axis.}
\label{fig:distribution of T}
\end{center}
\end{figure}

The reasoning above implies that the consensus occurs fast [i.e., $O(N)$ time]
or slowly [i.e., $O(N/C)$ time] with probability $1/2$ each. Consensus of the opposite states in the different cliques does not occur in the former case, and it occurs in the latter case.
To test this point, I carried out $10^4$ runs of the voter model
with $N=10^4$ and $M=1$.
Half the nodes in each clique was initially assumed to take the $\mathbf{0}$ state. The numerically obtained histogram of $T$ is shown in
Fig.~\ref{fig:distribution of T}.
The distribution of $T$ is in fact bimodal
with a heavy tail (note the logscale of the abscissa).
Each peak contains roughly half the runs.
In addition, the positions of the two peaks are 
roughly separated by $1/C=10^4$ times, which is consistent
with the ratio between the $O(N)$ and $O(N/C)$ consensus time estimated for each peak.

\section{Mean consensus time when $C<1$ via the coalescing random walk}\label{sec:calculation 6 configurations}

When $C<1$, the transition among configurations in the system of two coalescing random walkers
on the two-clique graph is described by
\begin{equation}
\bm p(t+1) = A \bm p(t)
\end{equation}
and
\begin{equation}
r(t+1) = \bm v \bm p(t),
\end{equation}
where
\begin{equation}
\bm p(t) = \left(p_1(t)\; p_2(t)\; \cdots\; p_6(t)\right)^{\top},
\end{equation}
$\top$ denotes the transposition,
\begin{align}
& A=\frac{1}{\Delta}\times \notag\\
&
\begin{pmatrix}
\Delta - 2NC-2 & N(1-C)-1 & 0 & 0 & 0 & 0\\
2NC & \Delta-N-1 & 1 & 0 & 2N(1-C) & 0\\
0 & 1 & \Delta-N-1 & 2NC & 0 & 2N(1-C)\\
0 & 0 & N(1-C) & \Delta-2NC & 0 & 0\\
0 & NC-1 & 0 & 0 & \Delta-2N(1-C)-4 & 2\left(1-\frac{1}{NC}\right)\\
0 & 0 & NC & 0 & 2 & \Delta-2N(1-C)-2
\end{pmatrix},
\end{align}
\begin{equation}
\bm v = \frac{1}{\Delta}
\left(2\quad 2\quad 0\quad 0\quad 2\quad \frac{2}{NC} \right),
\end{equation}
and
\begin{equation}
\Delta = 2N(N+C-1)
\end{equation}
is twice the number of links.

By adapting Eq.~\eqref{eq:<T> via coalescing RW large C} to the present system with six configurations, I obtain
\begin{equation}
r(t) = \bm x \bm p(0),
\end{equation}
where $\bm x = (x_1\; \cdots\; x_6)$ is the solution of
\begin{equation}
\bm x (I-A) = (1\; \cdots\; 1).
\end{equation}
In fact, I obtain
\begin{equation}
\bm x = c_0\begin{pmatrix}
2N (N^3C + 3N^2C + 3NC + 2C +1)C\\
2N^4C^2 + N^3(6C^2+C) + N^2(5C^2 +2C) + N(2C^2+3C) + 2C+1\\
N^4(2C^2+C) + N^3(7C^2+5C) + N^2(5C^2+5C) + N(2C^2+5C+1) + 2C+1\\
N^4(2C^2+C) + N^3(7C^2+6C) + N^2(6C^2+9C) + N(4C^2+10C+1) + 4C+2\\
2 N^2(N+1)(NC+2C+1)C\\
N \left[N^3(2C^2+C) + N^2(7C^2+4C) + N(4C^2+C) -2C-1\right]
\end{pmatrix},
\label{eq:x small}
\end{equation}
where
\begin{equation}
c_0 = \frac{N+C-1}{\left[N^3 C + N^2(C^2+4C)+ N(2C^2+5C)+2C+1\right]C}.
\label{eq:c_0 small}
\end{equation}
Given $C>1/N$ by definition, Eq.~\eqref{eq:c_0 small} implies $c_0 = O(1/N^2C^2)$ as $N\to\infty$. Therefore,
Eq.~\eqref{eq:x small} implies that $x_1, x_2, x_5 = O(N^2)$ and $x_3, x_4, x_6 = O(N^2/C)$. Therefore, $r(t)=O(N^2/C)$ in general. I implicitly normalized the time for the sake of the present analysis such that each node is updated once per time $2N$ on average. In terms of the rescaled time such that each node is updated once per unit time on average, I obtain $r(t)=O(N/C)$.

\newpage
\clearpage

\clearpage

\clearpage

\clearpage

\end{document}